\documentclass[final,12pt,5p,twocolumn]{elsarticle}

\usepackage{amssymb}

\usepackage{lineno}

\usepackage{hyperref}
\usepackage{flushend}
\newcommand{\figref}[1]{Fig.~\ref{#1}}
\newcommand{\eqref}[1]{Eq.~(\ref{#1})}
\newcommand{\secref}[1]{Sec.~\ref{#1}}
\newcommand{\cmsq}{$\mathrm{cm^2}$}
\newcommand{\megaohmsq}{$\mathrm{M \Omega/sq.}$}

\journal{Nuclear Instruments and Methods in Physics Research Section A}

\begin{document}

\begin{frontmatter}



\title{Development of ultra-low mass and high-rate capable RPC based on Diamond-Like Carbon electrodes for MEG~II experiment}


\author[u-tokyo]{Kensuke~Yamamoto\corref{cor}}
\ead{kensukey@icepp.s.u-tokyo.ac.jp}
\author[icepp]{Sei~Ban}
\author[icepp]{Kei~Ieki\fnref{fn1}}
\author[kobe]{Atsuhiko~Ochi}
\author[u-tokyo]{Rina~Onda}
\author[icepp]{Wataru~Ootani}
\author[u-tokyo]{Atsushi~Oya}
\author[kobe]{Masato~Takahashi}
\cortext[cor]{Corresponding author}
\fntext[fn1]{Current address: \it{ICRR, The University of Tokyo, 5-1-5 Kashiwa-no-Ha, Kashiwa, 277-8582, Japan}}

\affiliation[u-tokyo]{organization={Department of Physics, The University of Tokyo},
                     addressline={7-3-1 Hongo, Bunkyo-ku},
                     state={Tokyo},
                     postcode={113-0033},
                     country={Japan}}
\affiliation[icepp]{organization={ICEPP, The University of Tokyo},
                    addressline={7-3-1 Hongo, Bunkyo-ku},
                    state={Tokyo},
                    postcode={113-0033},
                    country={Japan}}
\affiliation[kobe]{organization={Department of Physics, Kobe University},
                   addressline={1-1 Rokkodai-cho, Nada-ku},
                   city={Kobe},
                   state={Hyogo},
                   postcode={657-8501},
                   country={Japan}}

\begin{abstract}
A new type of resistive plate chamber with thin-film electrodes based on diamond-like carbon is under development for background identification in the MEG~II experiment.
Installed in a low-momentum and high-intensity muon beam, the detector is required to have extremely low mass and a high rate capability.
A single-layer prototype detector with 2~cm $\times$ 2~cm size was constructed and evaluated to have a high rate capability of 1~MHz/\cmsq{} low-momentum muons.
For a higher rate capability and scalability of the detector size, the electrodes to supply high voltage were segmented at a 1 cm pitch by implementing a conductive pattern on diamond-like carbon.
Using the new electrodes, a four-layer prototype detector was constructed and evaluated to have a 46\% detection efficiency with only a single layer active at a rate of $\cal O$(10~kHz).
The result with the new electrodes is promising to achieve the required detection efficiency of 90\% at a rate of 4 MHz/\cmsq{} with all the layers active.
\end{abstract}




\end{frontmatter}


\section{Introduction}
\label{sec:introduction}
A novel background identification detector is under development for the MEG II experiment, aiming at a further sensitivity improvement in the $\mu \to e \gamma$ search \cite{megii-design}.
It detects a low-energy positron of 1--5~MeV from the radiative muon decay (RMD) with a high-energy $\gamma$-ray of $>$48~MeV detected by a $\gamma$ detector.
An extremely low-mass design of radiation length $X_0$ of 0.1\% and a high rate capability of up to 4~MHz/\cmsq{} are required since the detector is planned to be installed in a low-momentum (28~MeV/$c$) and high-intensity ($10^8$/s) muon beam.
The detection efficiency of 90\% for minimum ionising particle (MIP), the timing resolution of 1~ns, and the 20~cm diameter detector size are also required to efficiently identify the RMD.

\begin{figure}[tbp]
    \centering
    \includegraphics[width=8.5cm]{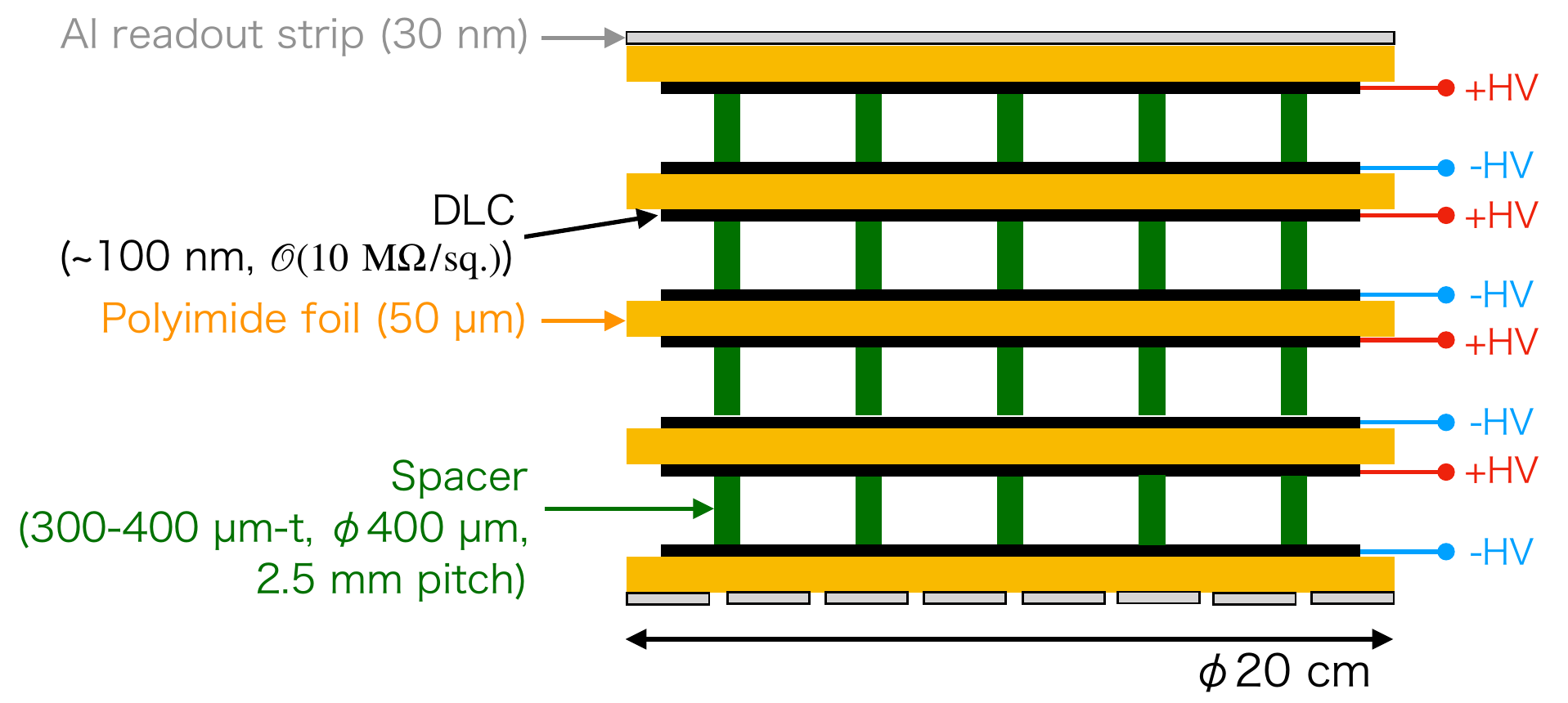}
    \caption{Concept of DLC-RPC}
    \label{fig:dlc-rpc-concept}
\end{figure}

A new type of resistive plate chamber (RPC) based on diamond-like carbon (DLC) electrodes (DLC-RPC) shown in \figref{fig:dlc-rpc-concept} is a promising candidate for the detector.
It has thin-film resistive electrodes based on DLC coating for the low-mass design.
The overall material budget can be suppressed to 0.095\% $X_0$ with a four-layer configuration.
A single-layer detection efficiency is required to be above 40\% to achieve the 90\% detection efficiency with four layers.

\section{Rate capability of DLC-RPC}
\label{sec:rate-capability}

The performance of RPCs generally degrades at a high rate due to voltage drop in the resistive electrode \cite{rpc-rate}.
In contrast to conventional glass-based RPCs, in which an avalanche current flows through the resistive plates perpendicularly, the current flows through the surface of the DLC layer transversely in the DLC-RPC.
Therefore, the further the current flow distance is, the larger the voltage drop is.
The voltage drop $\delta V$ for the DLC-RPC is given by
\begin{equation}
    \nabla^2 \delta V(x, y) = Q_{\mathrm{mean}}(V_{\mathrm{eff}}) \cdot f(x, y) \cdot \rho_{S},
    \label{eq:voltage-drop}
\end{equation}
where $Q_{\mathrm{mean}}(V_{\mathrm{eff}})$ is the mean of avalanche charge at an effective high voltage, $f(x, y)$ is the hit rate, and $\rho_{S}$ is the sheet resistivity of DLC.
The voltage drop is proportional to the square of the current flow distance approximately and to the sheet resistivity of DLC.
These parameters are important to design the resistive electrodes.

\section{Single-layer prototype}
\label{sec:single-layer-prototype}
A single-layer prototype detector with 2~cm $\times$ 2~cm size was constructed and its performance was evaluated \cite{tipp2021}.

\begin{figure}[tbp]
    \centering
    \includegraphics[width=8.8cm]{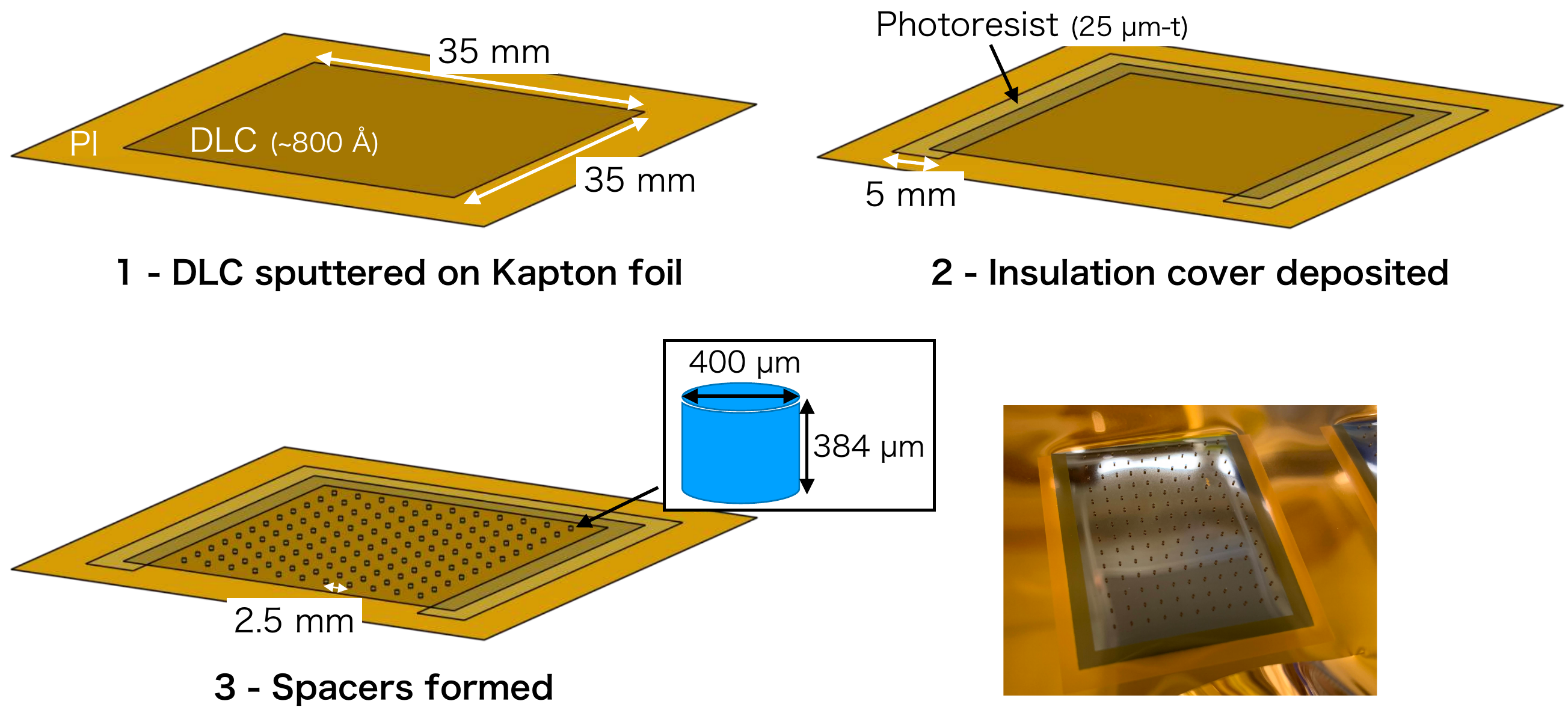}
    \caption{Electrode production processes and photo of an electrode for a single-layer prototype}
    \label{fig:first-electrode-prod}
\end{figure}

\subsection{Electrode production and construction}
\label{subsec:first-proto-construction}
The resistive electrode production uses sputtering and photolithographic technologies which are commonly used for micropattern gaseous detectors.
\figref{fig:first-electrode-prod} shows the electrode production processes.
Firstly, DLC was sputtered on 50~\textmu{}m-thick Kapton foil masked by a seal with a 3.5~cm square hole.
The sheet resistivity was 60~\megaohmsq{} and 7~\megaohmsq{} for the anode and cathode, respectively.
Secondly, 25~\textmu{}m-thick insulating photoresist film was deposited on the boundary of DLC to prevent discharge here.
Finally, photolithographic spacers with 400~\textmu{}m diameter and 384~\textmu{}m thickness were formed on DLC with 2.5~mm pitch.

\begin{figure}[tbp]
    \centering
    \includegraphics[width=7cm]{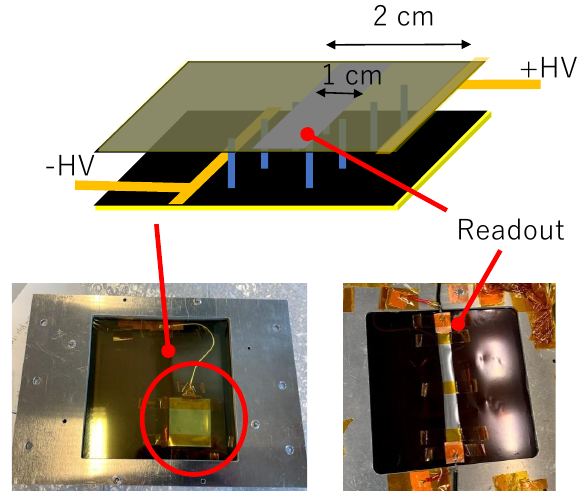}
    \caption{Constructed single-layer prototype \cite{tipp2021}}
    \label{fig:first-prototype}
\end{figure}

The construction scheme and setups are described in \cite{tipp2021}.
\figref{fig:first-prototype} shows the constructed single-layer prototype.
The current flow distance is 2~cm.
The induced signal in the readout strip was amplified by 38~dB amplifiers, and fed into DRS4 waveform digitiser \cite{drs}.

\subsection{Performance measurement at high rate}
\label{subsec:first-proto-test}

\begin{figure}[tbp]
    \centering
    \includegraphics[width=6cm]{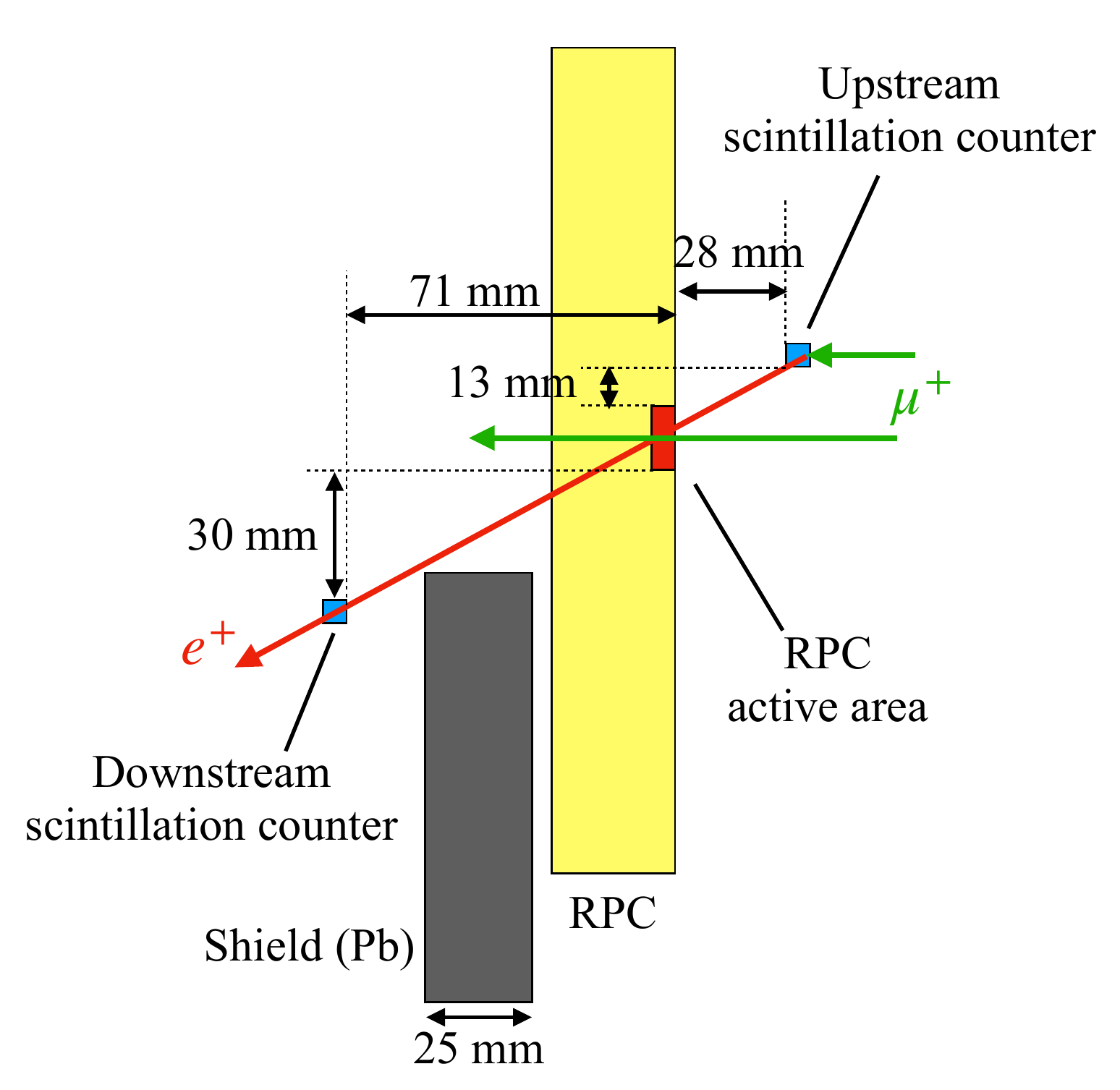}
    \caption{Setup of performance test at high rate}
    \label{fig:highrate-setup}
\end{figure}

The performance for the MIP positron detection in a low-momentum high-rate muon beam was measured at the $\pi$E5 beam line at Paul Scherrer Institute.
As shown in \figref{fig:highrate-setup}, the detector was exposed to a muon beam with a spread of ($\sigma_x$, $\sigma_y$) = (13~mm, 23~mm) at rates of 1~MHz/\cmsq{} and 3.5~MHz/\cmsq{}.
The readout region of the detector was aligned to the beam centre.
A part of the muons stopped in the upstream scintillation counter and decayed to positrons.
The decay positrons that passed through the detector were selected by the coincidence of upstream and downstream scintillation counters.
The angle between the muon beam axis and the triggered positron path was approximately 30$^{\circ}$, which is the typical incident angle of the RMD positrons in the MEG~II experiment.

\begin{figure}[tbp]
    \centering
    \includegraphics[width=8cm]{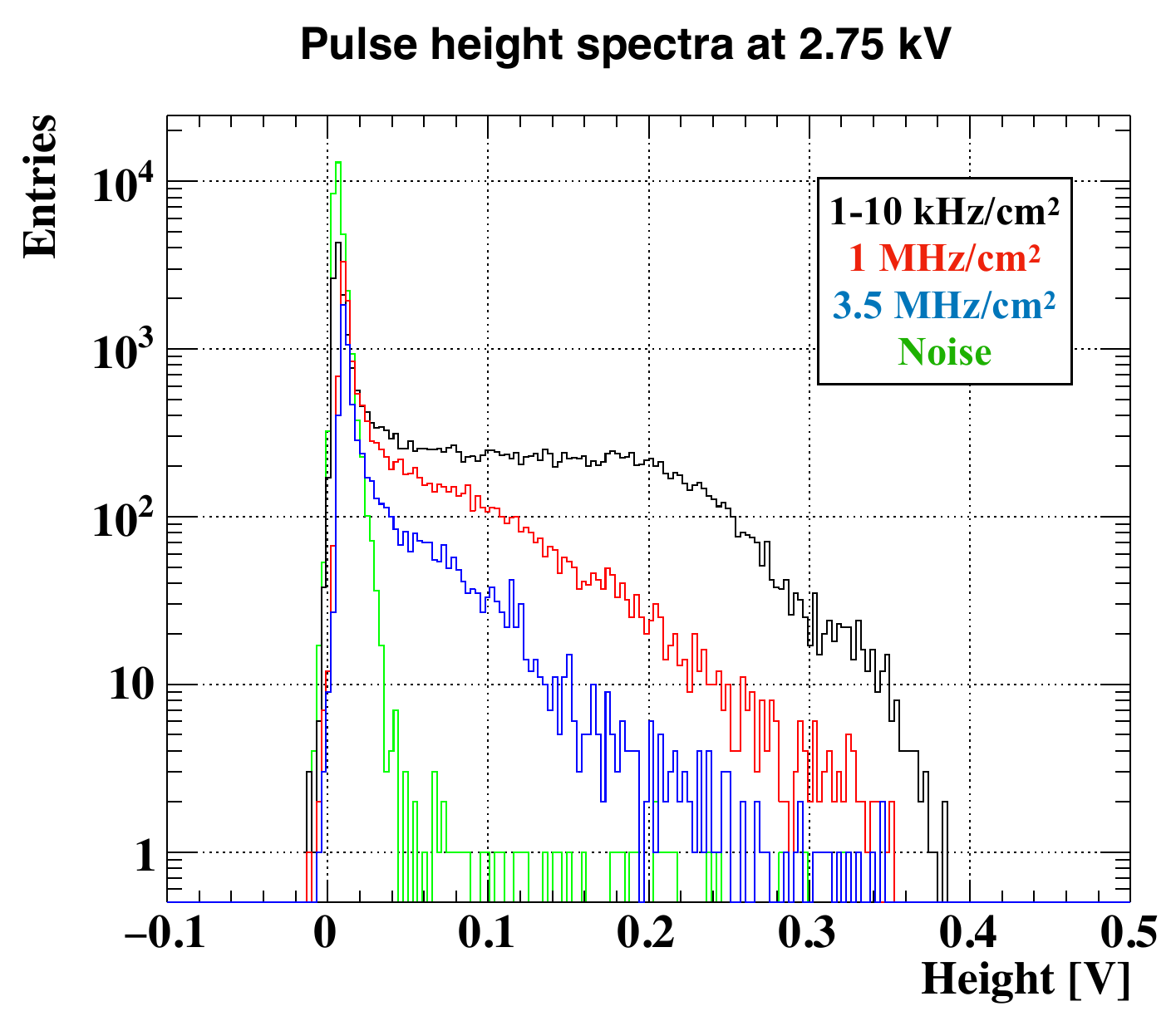}
    \caption{Pulse height spectra for MIP positron measured at different beam rates. The red spectrum is acquired in low-momentum muons at a rate of 1~MHz/\cmsq{}. The blue one is that at a rate of 3.5~MHz/\cmsq{}. The black one is that at a low rate of 1--10~kHz/\cmsq{} with another setup \cite{tipp2021}. The green one is a noise spectrum in off-timing at a low rate.}
    \label{fig:highrate-height-spectra}
\end{figure}

\figref{fig:highrate-height-spectra} shows the pulse height spectra for MIP positrons at applied high voltage of 2.75~kV after the selection from the hit time difference between scintillation counters to avoid accidentally triggered events.
The red and blue histograms are the spectra in low-momentum muons at 1~MHz/\cmsq{} and 3.5~MHz/\cmsq{}, respectively.
The black histogram is obtained at a low rate of 1--10~kHz/\cmsq{} with another setup \cite{tipp2021}.
The performance degradation was clearly observed with a higher beam rate.
A detection efficiency is defined as the fraction of the number of events with pulse height over 20~mV to the number of triggered events.
The threshold of 20~mV was set according to the noise level shown in a green histogram in \figref{fig:highrate-height-spectra}.
The fraction of noise contamination in counting signal events at the 20~mV threshold was less than 2\%. 
A 42\% (17\%) detection efficiency was achieved at 1~MHz/\cmsq{} (3.5~MHz/\cmsq{}).

The voltage drop was calculated to 110--170~V and 210--310~V at 1~MHz/\cmsq{} and 3.5~MHz/\cmsq{}, respectively, using \eqref{eq:voltage-drop}, where $Q_{\mathrm{mean}} = 2.3~\mathrm{pC}$ for 1~MHz/\cmsq{} and $Q_{\mathrm{mean}} = 1.2~\mathrm{pC}$ for 3.5~MHz/\cmsq{}.
These voltage drops are consistent at 10\% with the observed detection efficiencies.
It is concluded that the single-layer prototype has a rate capability of up to 1~MHz/\cmsq{} low-momentum muons.

\section{Four-layer prototype}
\label{sec:four-layer-prototype}
The single-layer prototype did not still satisfy the requirement of the rate capability.
The voltage drop should be suppressed for a higher rate capability.
As described in \secref{sec:rate-capability}, it is necessary to shorten the current flow distance and to make the sheet resistivity of DLC lower.

Segmented high voltage supply with 1 cm pitch and a low DLC resistivity of 10~\megaohmsq{} were designed to fulfill the rate capability requirement of 4~MHz/\cmsq{}.
The segmentation is also necessary for the scalability of the detector size.
The voltage drop can be suppressed to 60--80~V at 4~MHz/\cmsq{} with these configurations. 

\subsection{Electrode production}
\label{subsec:improved-proto-electrode}

\begin{figure}[tbp]
    \centering
    \includegraphics[width=8.8cm]{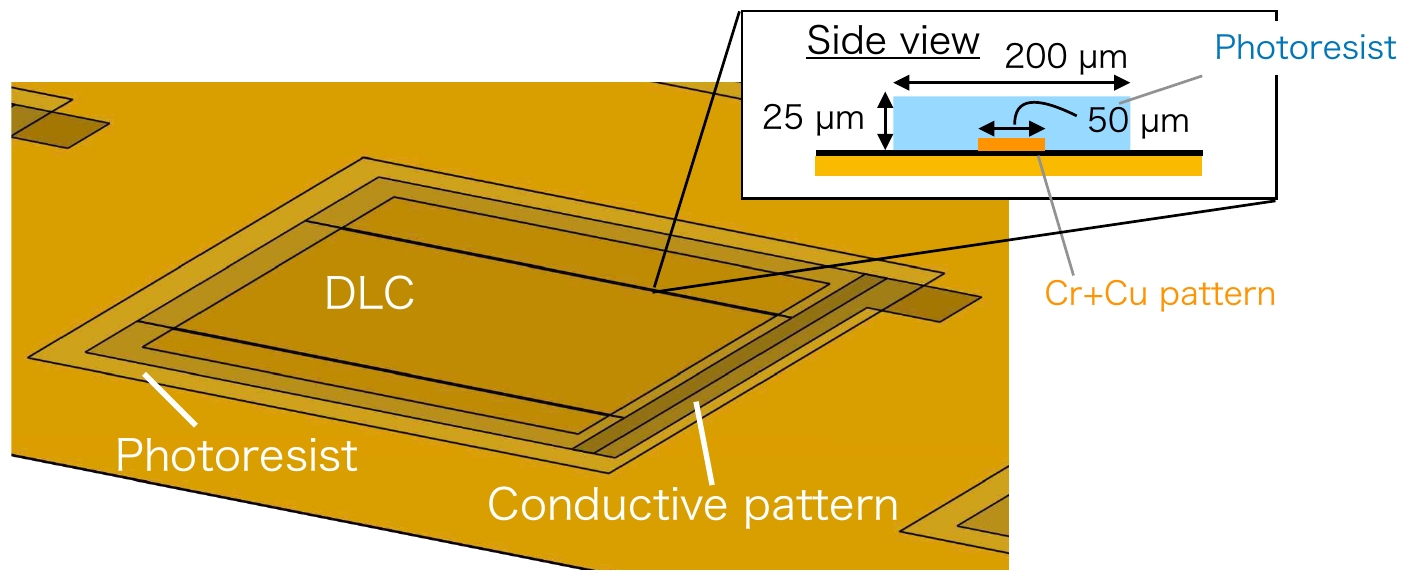}
    \caption{Conductive and insulation pattern on DLC for high voltage supply segmentation}
    \label{fig:conductive-pattern}
\end{figure}

The 3~cm $\times$ 3~cm electrodes with the segmented high voltage supply were produced.
Firstly, DLC was sputtered as in the electrodes for the single-layer prototype.
The sheet resistivity was 20~\megaohmsq{}
In order to segment the high voltage supply, a conductive pattern which consists of 50~\AA{}-thick chromium layer and 2500~\AA{}-thick copper layer was formed on the DLC by the liftoff method as shown in \figref{fig:conductive-pattern}.
The conductive pattern was covered with 25~\textmu{}m-thick insulating photoresist to avoid discharge on it.
The widths of the conductive strip and of the insulation strip were 50~\textmu{}m and 200~\textmu{}m, respectively.
The region on the insulation strip will be inactive due to the charge-up effect.
Implementation of conductive and insulation strips will increase the inactive area by 2\%.
The quality of electric connection between the conductive pattern and the DLC was confirmed to be connected well by repeating measurements of the resistance between the conductive pattern and a point in DLC several times.

\begin{figure}[tbp]
    \centering
    \includegraphics[width=7cm]{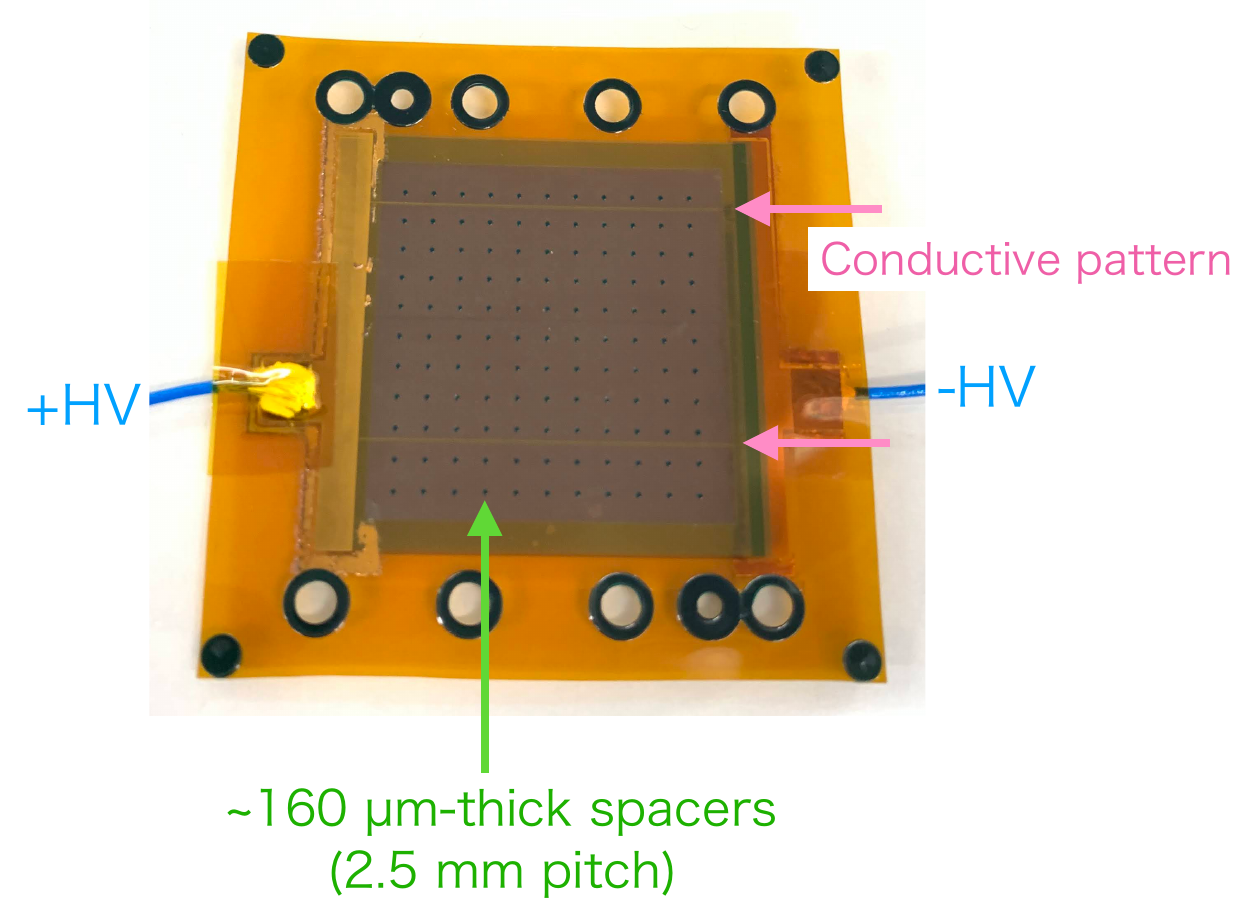}
    \caption{Picture of produced electrode}
    \label{fig:improved-electrode}
\end{figure}

Because the production of photoresist used for the spacers of the single-layer prototype was cancelled, solder resist had to be used instead.
The problem was that 300~\textmu{}m-thick spacers cannot be formed using the solder resist.
As an alternative solution, 160~\textmu{}m-thick spacers were attached on both sides of the electrode and accumulated with precise alignment to form enough thick gaps.
\figref{fig:improved-electrode} shows a picture of the produced electrode.

The outermost resistive electrodes were made of films which were aluminised on the outer face and DLC-coated on the inner face.
Aluminium readout strips were formed by etching.

\subsection{Construction}
\label{subsec:improved-proto-construction}

\begin{figure}[tbp]
    \centering
    \includegraphics[width=9cm]{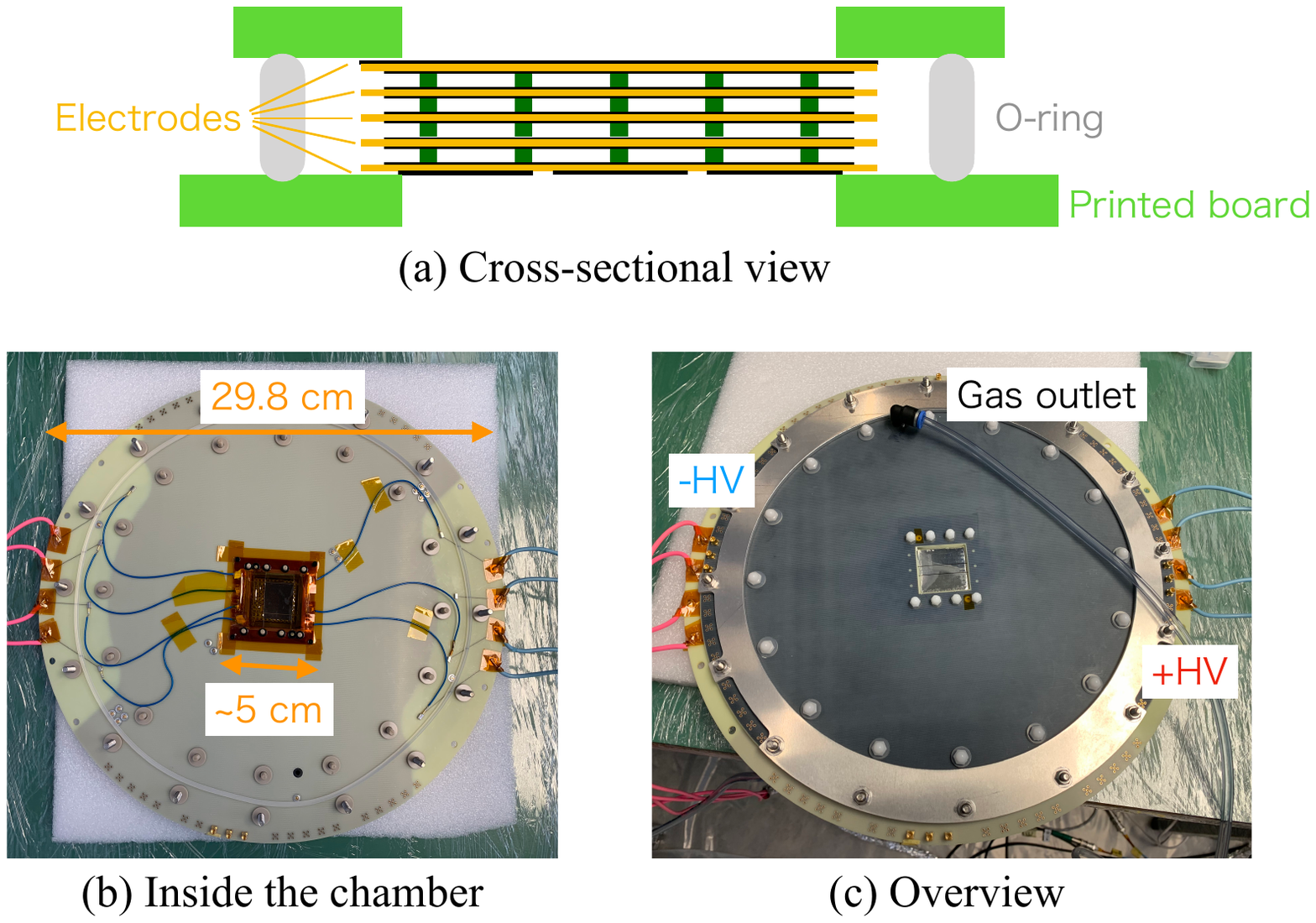}
    \caption{Constructed four-layer prototype}
    \label{fig:module0}
\end{figure}

\figref{fig:module0} shows the assembled four-layer prototype.
The outermost electrodes were glued onto printed boards.
The intermediate electrodes with resistive layers on both sides were stacked along alignment pins.
It was filled with a gas mixture of R134a(94\%) + SF$_6$(1\%) + isobutane(5\%).
The high voltage was supplied by conductive glue on the conductive pattern.
The signal was read out from both ends of the aluminium strips, amplified by 38~dB amplifiers, and fed into the DRS4 waveform digitiser \cite{drs}.

\subsection{Performance test}
\label{subsec:improved-proto-test}

High voltage can be applied to each layer independently since electrodes are decoupled by the polyimide film.
Thus, the detection efficiency to $\beta$-ray at ${\cal O}$(10~kHz) was measured for each layer independently.

An issue of quality control of gas gaps was found due to a positive pressure inside the chamber.
This caused a smaller gas gain at the centre than that at the edge since the electrode films expanded outwards.
In addition, it caused unstable operation in three layers of four.
A 46\% detection efficiency was measured using the edge of the active region in the second layer from the top in \figref{fig:module0}~(a), which was the only layer successfully operated.
It means that there are no problems with electrode design and detector system in principle.
The issue can be solved by negative pressure in the gas gap.
Further studies are planned to evaluate the performance of the four-layer prototype.

\section{Conclusion}
\label{sec:conclusion}
A new type of RPC with DLC electrodes is under development for RMD $\gamma$-ray identification in the MEG~II experiment.
Using the single-layer prototype with $2~\mathrm{cm} \times 2~\mathrm{cm}$ size, the high rate capability of 1~MHz/\cmsq{} was achieved.
The resistive electrode was improved by the conductive pattern on DLC for a higher rate capability of up to 4~MHz/\cmsq{} and for its scalability.
The four-layer prototype with improved electrodes was tested, and promising results were obtained although there was an issue of quality control of the gas gaps.
Further performance measurements with $\beta$-ray at ${\cal O}$(100~kHz) and with high-intensity muon beam at up to 8~MHz/\cmsq{} are planned, which will be presented in the near future.

\section*{Acknowledgement}
\label{sec:acknowledgement}
I would like to thank H. Uehara of TRENG-F Products, Inc., whose great efforts contributed to the production of the improved electrodes.
This work is supported by MEXT/JSPS KAKENHI Grant Number 21H04991, JSPS Core-to-Core Program, A. Advanced Research Networks JPJSCCA20180004, and JSPS Overseas Challenge Program for Young Researchers 202280015.



 \bibliographystyle{elsarticle-num} 
 \bibliography{cas-refs}





\end{document}